\documentclass[a4paper,fleqn]{cas-dc}

\usepackage[numbers]{natbib}
\def\tsc#1{\csdef{#1}{\textsc{\lowercase{#1}}\xspace}}
\tsc{WGM}
\tsc{QE}
\tsc{EP}
\tsc{PMS}
\tsc{BEC}
\tsc{DE}

\begin{document}
\let\WriteBookmarks\relax
\def\floatpagepagefraction{1}
\def\textpagefraction{.001}
\let\printorcid\relax

\shorttitle{Plug-and-Play Convolutions with Topology Constraints}
\shortauthors{Xiao Zhang et~al.}

\title [mode = title]{PASC-Net:Plug-and-play Shape Self-learning Convolutions Network with Hierarchical Topology Constraints for Vessel Segmentation}                      



\author[1]{Xiao Zhang}

\fnmark[1]

\author[1]{Zhuo Jin}
\fnmark[1]

\author[1]{Shaoxuan Wu}

\author[1]{Fengyu Wang}

\author[1]{Guansheng Peng}

\author[1]{Xiang Zhang}
\cormark[1]
\ead{xiangz@nwu.edu.cn}

\author[1]{Ying Huang}

\author[2]{JingKun Chen}

\author[1]{Jun Feng}
\cormark[1]
\ead{fengjun@nwu.edu.cn}

\affiliation[1]{organization={School of Information Science and Technology, Northwest University},
                city={Xi'an},
                country={China}}

\affiliation[2]{organization={Institute of Biomedical Engineering, Department of Engineering Science, University of Oxford},
                city={Oxford},
                country={UK}}
\nonumnote{This work was supported by the National Natural Science Foundation of China (No.62403380, 62401468). }
\cortext[cor1]{Corresponding author}
\fntext[fn1]{Co-first author.}

\begin{abstract}
Accurate vessel segmentation is crucial to assist in clinical diagnosis by medical experts. However, the intricate tree-like tubular structure of blood vessels poses significant challenges for existing segmentation algorithms. Small vascular branches are often overlooked due to their low contrast compared to surrounding tissues, leading to incomplete vessel segmentation.
Furthermore, the complex vascular topology prevents the model from accurately capturing and reconstructing vascular structure, resulting in incorrect topology, such as breakpoints at the bifurcation of the vascular tree.
To overcome these challenges, we propose a novel vessel segmentation framework called PASC-Net. It includes two key modules: a plug-and-play shape self-learning convolutional (SSL) module that optimizes convolution kernel design, and a hierarchical topological constraint (HTC) module that ensures vascular connectivity through topological constraints. 
Specifically, the SSL module enhances adaptability to vascular structures by optimizing conventional convolutions into learnable strip convolutions, which improves the network's ability to perceive fine-grained features of tubular anatomies. Furthermore, to better preserve the coherence and integrity of vascular topology, the HTC module incorporates hierarchical topological constraints—spanning linear, planar, and volumetric levels—which serve to regularize the network's representation of vascular continuity and structural consistency.
We replaced the standard convolutional layers in U-Net, FCN, U-Mamba, and nnUNet with SSL convolutions, leading to consistent performance improvements across all architectures. Furthermore, when integrated into the nnU-Net framework, our method outperformed other methods on multiple metrics, achieving state-of-the-art vascular segmentation performance. The code is available at \href{https://github.com/IPMI-NWU/PASC-Net}{https://github.com/IPMI-NWU/PASC-Net}.
\end{abstract}




\maketitle

\section{Introduction}
Vascular segmentation is a critical task in medical image analysis, underpinning a wide range of clinical applications such as the diagnosis and monitoring of cardiovascular diseases (CVDs), detection of tumors through angiogenesis assessment, and early screening of retinal disorders like diabetic retinopathy\cite{amin2020distinctive,de2018clinically,gaziano2006cardiovascular,moccia2018blood}. 
High-quality vessel segmentation enables clinicians to quantify morphological features such as vessel width, tortuosity, and branching patterns, which are key biomarkers for disease progression and treatment planning\cite{guerrero2007real,laibacher2019m2u,zhang2022progressive,zhang2023anatomy}. As imaging modalities like Coronary Angiography (CAG) and fundus photography continue to evolve in resolution and accessibility, the demand for automated, accurate, and robust vessel segmentation algorithms has grown substantially. 
Therefore, the development of effective vessel segmentation techniques is both of academic interest and critical practical importance in modern precision medicine.
\begin{figure}[ht]
    \centering
    \includegraphics[width=0.43\textwidth]{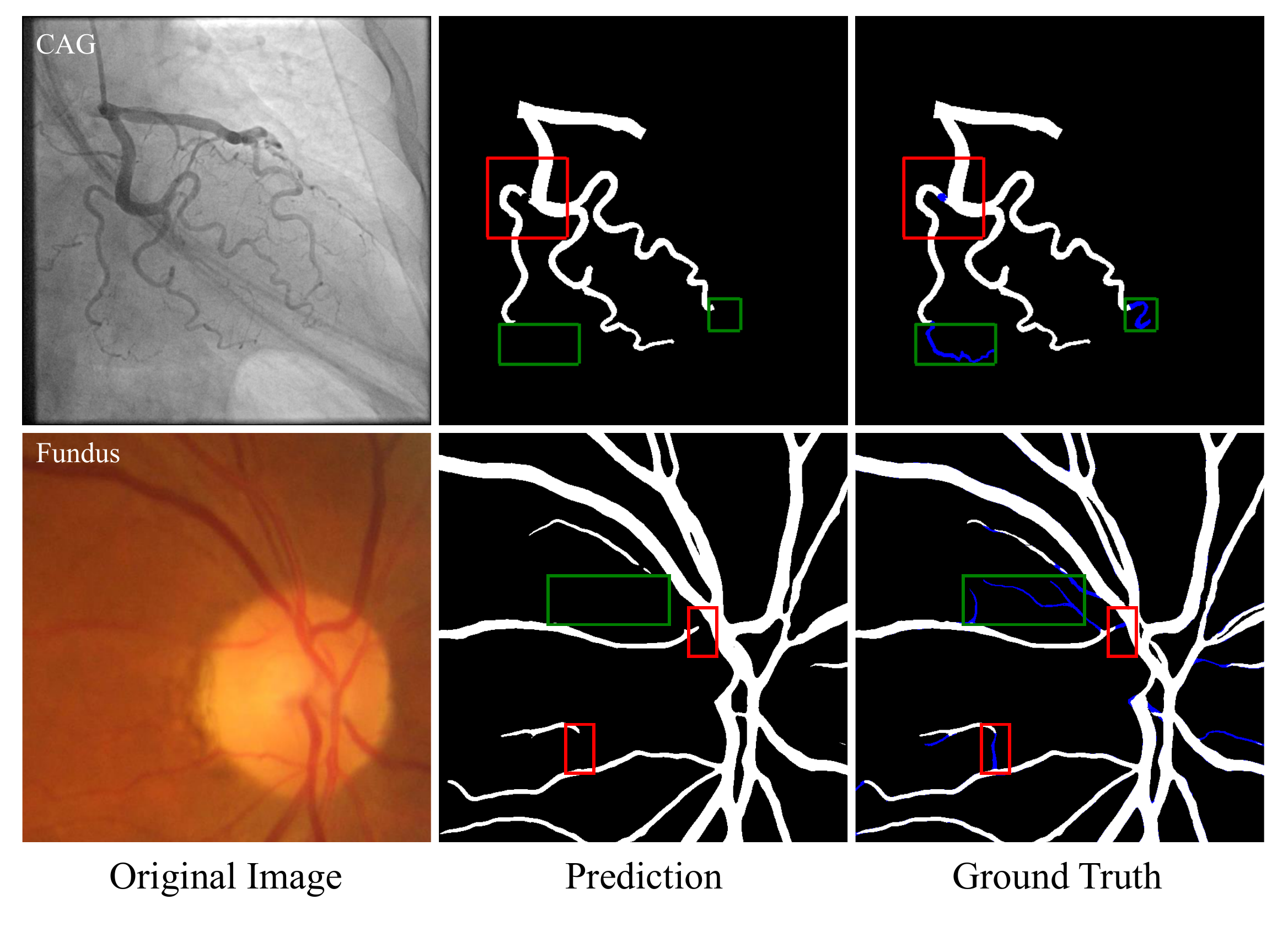}
    \caption{The challenges of vessels segmentation: 1) Lost small branches (green box): Small branches at the end of the vascular tree are lost due to low contrast; 2) Incorrect topological structures (red box): The appearance of breakpoints disrupts the vascular topology. The left side shows the original image, the middle displays the prediction results of nnUNet, and the right side shows the ground truth. }
    \label{fig:intro}
    \vspace{-20pt}
\end{figure}
Although computer-aided automatic vascular segmentation has substantially improved diagnostic efficiency and reliability, several critical challenges persist, as illustrated in Fig.\ref{fig:intro}. 
\textbf{First, the loss of small branches.} The diameter of blood vessels varies significantly and exhibits diverse morphologies in images. For example, in CAG images, the diameter of coronary arteries ranges from approximately 1.8 mm to 7.3 mm\cite{diameter}. This characteristic makes it challenging for models to accurately detect fine vessels, as shown in the green box in Fig. \ref{fig:intro}, often resulting in missed detections.
\textbf{Second, incorrect topological structures.} In coronary and retinal images, vessels exhibit a tree-like tubular structure with complex bifurcations, directional changes, and regional heterogeneity. This highly intricate spatial relationship makes it difficult for models to learn complete connectivity, particularly leading to fractures at critical junctions and branch points, thereby disrupting the overall topological structure. 
These challenges highlight the need for a dedicated segmentation framework that can effectively capture fine-scale vascular features while preserving global anatomical topology.

To address the challenge of small vessel branch loss, CNN-based methods \cite{yan2018three,chen2021retinal,yan2018joint,atli2021sine,livne2019u} have significantly improved vascular segmentation performance through local feature extraction capabilities, mitigating the reliance on handcrafted features in traditional approaches\cite{moccia2018blood,frangi1998multiscale}. However, due to their limited receptive fields, these methods still fail to effectively model long-range dependencies, resulting in missed detection of fine vascular branches. The introduction of Transformer models \cite{alexey2020image,chen2021transunet} established global dependencies through self-attention mechanisms, partially resolving the issue of small vessel continuity recognition. Yet their high computational complexity and slow inference speed limit practical clinical applicability. The recently proposed Mamba architecture \cite{ma2024u} enhances computational efficiency while maintaining long-range dependency modeling capabilities, but its preference for global features often comes at the expense of local details, leading to incomplete segmentation of fine vessels in low-contrast regions.

To tackle the challenge of erroneous vascular topology, DSCNet \cite{qi2023dynamic} partially improved vascular connectivity by incorporating topological priors such as dynamic shape convolution, while CoANet \cite{mei2021coanet} employed strip-shaped convolution to simulate tubular structures, enhancing topological learning capability. However, these methods rely on idealized assumptions about vascular spatial distribution, making them difficult to adapt to the complex geometric variations of real-world vasculature. Existing approaches have yet to simultaneously maintain fine vessel integrity and ensure topological accuracy, urgently requiring a unified framework capable of synergistically addressing these two critical issues.

Based on these challenges, we propose a novel vascular segmentation framework named PASC-Net, which incorporates a plug-and-play shape self-learning convolutional (SSL) module and hierarchical topology constraint (HTC) module. 
The SSL module enhances vascular spatial representation through the design of four self-learning shaped strip convolutions, enabling seamless integration into diverse convolutional network architectures. 
The HTC module is designed to address the issue of incorrect topology by preserving the coherence and integrity of vascular topology from three levels: line, surface, and volume. 
To validate the effectiveness of our method, we conducted extensive experiments on two datasets and evaluated the proposed two modules across multiple network frameworks, demonstrating their effectiveness and good portability. The contribution of our work can be summarized as follows: 
\begin{itemize}
\item[$\bullet$] The proposed plug-and-play SSL module, which incorporates four direction-aware strip-shaped convolutional kernels to enhance the network's capability in modeling thin and morphologically varied vascular structures.
\item[$\bullet$] The proposed HTC module comprehensively constrains vascular structures across three hierarchical levels—line, surface, and volume dimensions—by calculating centerlines, vessel masks, and 8-connected neighborhood matrices. This approach addresses issues such as discontinuities and topological errors, significantly enhancing the coherence of vascular structures.
\item[$\bullet$] Experimental results on two public datasets demonstrate that the proposed framework significantly outperforms existing state-of-the-art methods. Furthermore, systematic evaluations across multiple network architectures further validate the effectiveness and strong transferability of the SSL and HTC modules.
\end{itemize}

\section{Related Works}
Traditional methods for vascular segmentation include threshold-based approaches, edge detection, morphological operations, and image segmentation\cite{moccia2018blood,kirbas2004review}. While these methods perform well in handling simple vascular structures, they often fall short in complex vascular networks. In recent years, the continuous advancements in deep learning technology have led to the emergence of innovative algorithms in the field of vascular segmentation. Deep learning-based vascular segmentation methods can be categorized into two main types: 1) Methods based on convolutional neural networks and transformer models; 2) Methods based on topological constraints.

\begin{figure*}[ht]
  \centering
  \includegraphics[width=\textwidth]{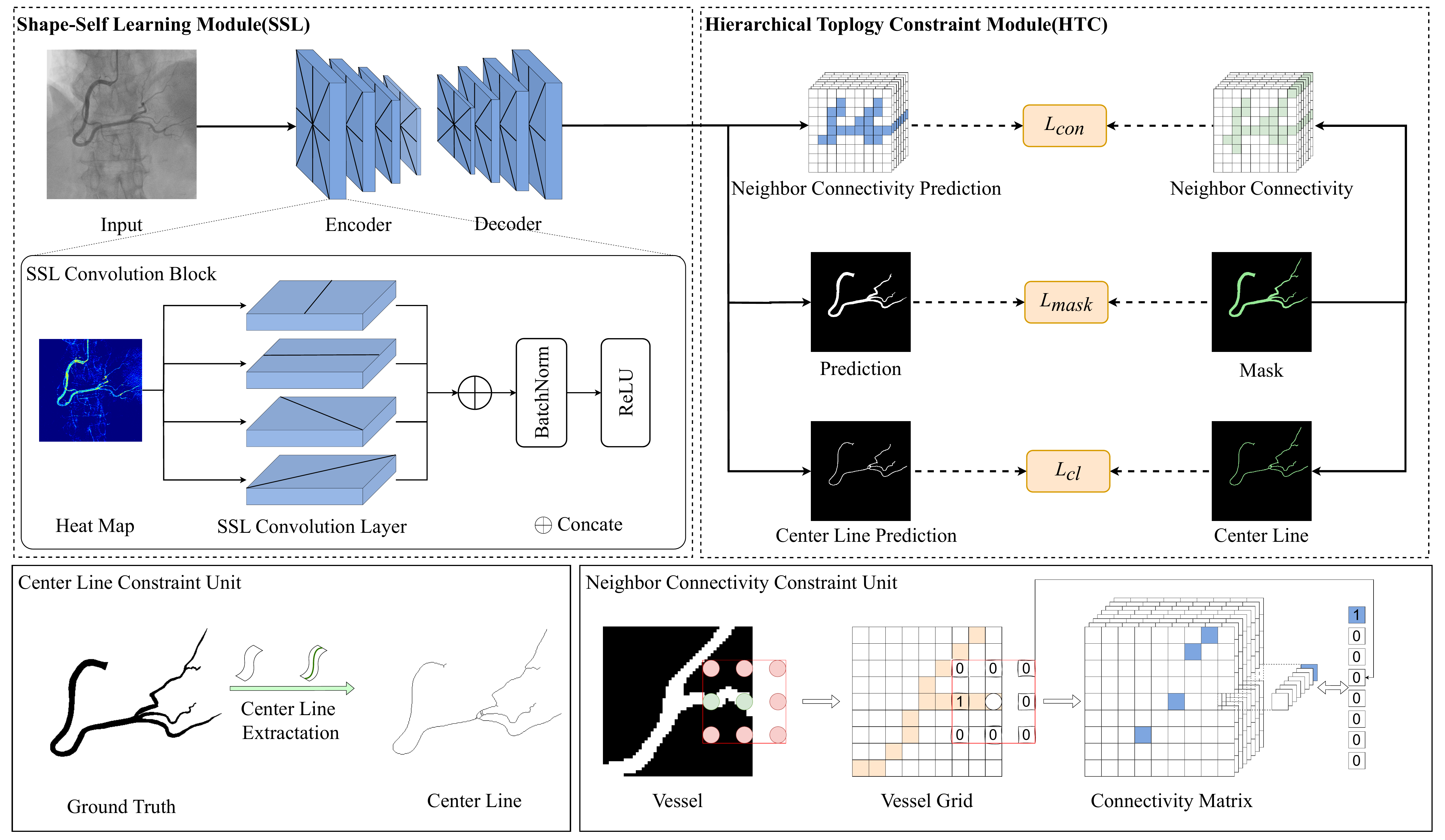}  
  \caption{The proposed PASC-Net framework, including overall network architecture, SSH module, HTC module (Neighbor Connectivity Constraint Unit and Center Line Constraint Unit).}
  \label{fig:spect}
  \vspace{-20pt}
\end{figure*}

\subsection{Methods based on CNN and Transformer}
Convolutional neural networks (CNNs) have been the cornerstone of medical image segmentation, with various architectures achieving remarkable performance\cite{wu2022vessel,liming2023adaptive,holroyd2023tube,lin2022ds,pan2023eg,park2022tcu}. Among them, U-Net \cite{ronneberger2015u} has become a standard in the field, particularly for medical images. Fully Convolutional Networks (FCNs) further advanced segmentation by allowing pixel-wise prediction without fully connected layers, offering more efficient models for tasks with dense outputs\cite{long2015fully}.

Building upon these foundational CNN architectures, newer models such as ResUNet \cite{xiao2018weighted} and nnUNet \cite{isensee2021nnu} have emerged. ResUNet integrates residual connections into the U-Net architecture, enabling more efficient feature propagation and better gradient flow, which helps improve segmentation accuracy by addressing the vanishing gradient problem in deeper networks. On the other hand, nnUNet represents a self-adapting framework designed to optimize the U-Net architecture for different datasets and tasks. It automates several crucial aspects, such as preprocessing, augmentation, and network configuration, tailoring the architecture specifically for a given dataset and substantially improving performance without manual tuning.
Another notable model is U-Mamba \cite{ma2024u}, which combines the strengths of U-Net and Mamba networks. U-Mamba is designed to handle the challenge of long-range dependency loss, which is particularly significant in vascular segmentation tasks where vessels span large areas. 

However, these methods do not leverage the topological structure of blood vessels, often resulting in incorrect topological results. Besides, many of these methods feature complex architectures that require significant computational resources.

\subsection{Methods based on topological constraints}
Several studies have introduced topological constraints and specific loss functions to improve the connectivity and topological conherence of segmentation results\cite{he2024retinal,kande2023msr,jian2023dual,jia2023boundary,rouge2023cascaded,nader2023using,shi2023affinity,qiu2023corsegrec,tan2023multi,shi2023freecos,zhang2022progressive}. Shit et al. \cite{shit2021cldice} proposed clDice, a novel loss function designed to ensure topological correctness by utilizing vessel centerlines as a constraint. This approach effectively addresses the challenge of disconnected segmentations, a common issue in vascular imaging, by encouraging alignment between predicted and true centerlines. 
Qi et al. \cite{qi2023dynamic} introduced DSCNet (Dynamic Snake Convolution Network), which focuses on capturing tubular structures in vascular networks. By leveraging a dynamic snake convolution technique, DSCNet adapts convolutional filters to the shape of the vessels, improving the network’s ability to capture tubular structures. 
Yao et al. \cite{mei2021coanet} proposed CoANet, a method that integrates one-dimention convolution to enhance the network's sensitivity to road-like, linear features, such as those found in vascular structures. CoANet introduces a novel directional convolution mechanism that prioritizes feature extraction along vessel-like paths, improving the detection and connectivity of tubular structures, crucial for accurate vascular segmentation.

Although these methods use different constraints for topological enhancement, the constraints used are relatively simple. The comprehensive topology of blood vessels is not taken into account, leading to topological consistency of blood vessels cannot be comprehensively improved.

\section{Method}
As shown in Fig.\ref{fig:spect}, our proposed PASC-Net framework consists of two key modules. The first module is the shape self-learning convolutional (SSL) module, which enhances the ability to represent tree-like tubular structure features by introducing dynamically adjustable convolution kernels. This enables the capture of fine-grained features of vascular structures. The second is the hierarchical topology constraint (HTC) module, which further preserves the consistency of the vascular topology by establishing hierarchical topology constraints on lines, surfaces, and volumes.
\begin{figure*}[ht]
  \centering
  \includegraphics[width=\textwidth]{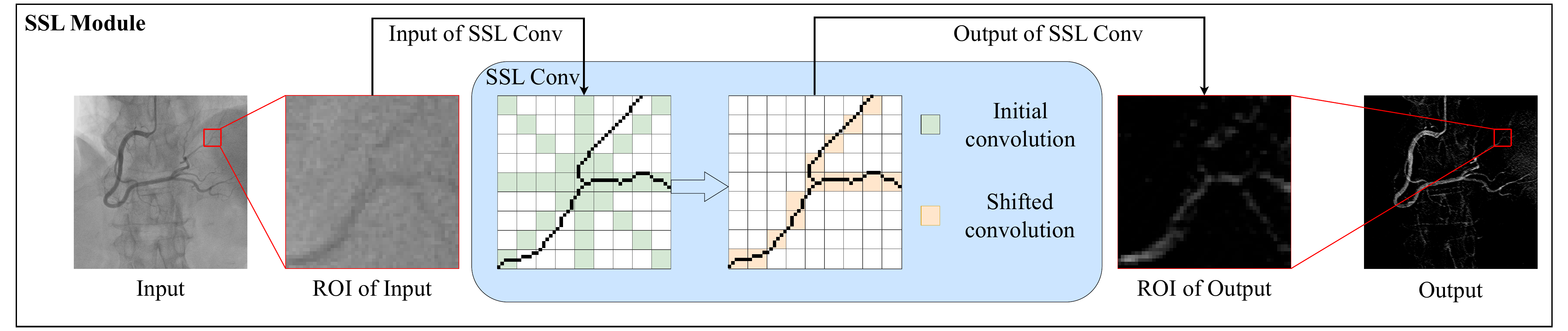}  
  \caption{The detail process of SSL Module. SSL Convolution slides over the input feature map to calculate the output, and automatically changes shape according to the shape of the blood vessel during the sliding process.}
  \label{fig:sslconv}
  \vspace{-20pt}
\end{figure*}

\subsection{Shape Self-Learning Module}
Since the advent of Convolutional Neural Networks (CNNs), standard convolutional kernels have been widely used for feature extraction and have performed well in most tasks. However, for vessels, which are tree-like tubular structures, standard square convolutional kernels have limitations in capturing their features, particularly in detecting small branches of vessels. To overcome this challenge, we designed a convolutional kernel that can automatically learn the tubular structure of vessels, as shown in Fig.\ref{fig:sslconv}.

The learning of SSL convolution kernel occurs in two aspects: on one hand, the convolutional kernel adaptively learns the tubular structure features of the vessels during feature extraction; on the other hand, the shape of the convolutional kernel is dynamically adjusted according to the shape of the vessels. Specifically, the initial SSL convolutional kernel consists of strip convolutions in four directions (horizontal, vertical, diagonal, and anti-diagonal), denoted as \(x\), \(y\), \(z\), and \(w\), respectively. In our experiments, the length of the strip convolution was set to 9, which not only corresponds to the standard $3\times3$ convolution but also fully represents the structure of the vessels, avoiding the issue of overly short kernels affecting performance.
Due to our SSL convolution kernel consists of four strip convolutions, the result of the SSL convolution operation is the sum of the results of four simple sub-operations. Different from ordinary convolution, SSL convolution kernel introduces an offset operation before calculating. Therefore, the convolution process in the SSL module can be divided into an offset process and a calculation process.
\subsubsection{Offset process}
After the initialization of four strip convolutions, the network begins to learn the tubular structure features of the vessels. Specifically, for a strip convolution of length m, fixing its midpoint, while the remaining m-1 points undergo perpendicular shifts to adapt to and learn the tubular structure of the vessels.
The computation process is as follows: For a convolution kernel \(W\) of length \(m\), let \(W_i\) represent the \(i\)-th point of strip convolution, and let  \(\delta_i\) represent the shift of the \(i\)-th point. The shift direction of each point in the convolutional kernel is perpendicular to the initial convolution, and it can shift either upward (\(\delta_i\) > 0) or downward (\(\delta_i\) < 0). To ensure that the shifted convolution maintains the continuity of the vessel, we employ an iterative shifting strategy. The shift of the midpoint (\(\delta_{m/2}\)) is set to 0, and the shifts of other points are based on the neighboring point closed to the midpoint. Thus, the relative shift between any two neighboring points is kept no more than 1 to ensure that the shape of the convolutional kernel remains smooth and continuous. After the offset process, initial convolution \(x\), \(y\), \(z\), and \(w\) turns into \(x'\), \(y'\), \(z'\), and \(w'\).

\subsubsection{Calculation process}
After the offset process, SSL Convolution begins the convolution calculation operation with feature maps, similar to standard convolution.
Given an input feature map \( X \), let \( Y_1, Y_2, Y_3, Y_4 \) represent the outputs of the \( x' \), \( y' \), \( z' \), and \( w' \) convolutions, respectively. And let \( W \) represent the weights of the convolutional kernel. Furthermore, based on the convolutional invariance theorem, the shifts in the convolutional kernel can be equivalently transformed into shifts in the feature map, thereby \( Y_1, Y_2, Y_3, Y_4 \) can be expressed as follows:
\begin{equation}
\begin{aligned}
Y_1(i,j) &= \sum_{m=-n}^{n} X(i{+}\Delta x,j{+}m) W(i,j{+}m) \\
          &= \sum_{m=-n}^{n} X(i,j{+}m) W(i{+}\Delta x,j{+}m),
\end{aligned}
\end{equation}

\begin{equation}
\begin{aligned}
Y_2(i,j) &= \sum_{m=-n}^{n} X(i{+}m,j{+}\Delta y) W(i{+}m,j) \\
          &= \sum_{m=-n}^{n} X(i{+}m,j) W(i{+}m,j{+}\Delta y),
\end{aligned}
\end{equation}

\begin{equation}
\begin{aligned}
Y_3(i,j) &= \sum_{m=-n}^{n} X(i{+}m{+}\Delta x,j{+}m{+}\Delta y) W(i{+}m,j{+}m)\\
&= \sum_{m=-n}^{n} X(i{+}m,j{+}m) W(i{+}m{+}\Delta x,j{+}m{+}\Delta y),
\end{aligned}
\end{equation}

\begin{equation}
\begin{aligned}
Y_4(i,j) &= \sum_{m=-n}^{n} X(i{-}m{+}\Delta x,j{+}m{+}\Delta y) W(i{+}m,j{+}m)\\
&= \sum_{m=-n}^{n} X(i{-}m,j{+}m) W(i{+}m{+}\Delta x,j{+}m{+}\Delta y),
\end{aligned}
\end{equation}
where \( \Delta x_m \) and \( \Delta y_m \) represent the shifts in the horizontal and vertical directions, respectively. The final SSL convolution result is the sum of these:
\begin{equation} 
Y(i, j) = Y_1(i, j) + Y_2(i, j)+ Y_3(i, j)+ Y_4(i, j). 
\end{equation}



where \( \Delta x_m \) and \( \Delta y_m \) represent the shifts in the horizontal and vertical directions, respectively. The final SSL convolution result is the sum of these:
\begin{equation} Y(i, j) = Y_1(i, j) + Y_2(i, j)+ Y_3(i, j)+ Y_4(i, j). \end{equation}

\subsection{Hierarchical Topological Constraints}
In order to alleviate the problem of incorrect topology caused by breakpoints in the bifurcation results of blood vessel segmentation, we propose a hierarchical topology constraint module. The constraints preserve the integrity and coherence of the vascular topology at three levels: line, surface, and volume. By employing a hierarchical approach, we ensure a more coherent segmentation, effectively preserving the integrity of the vascular network. HTC module not only enhances segmentation accuracy but also provides a robust framework for modeling the hierarchical relationships inherent in vascular topology structures.

\subsubsection{Centerline Constraint}  
The centerline derived from segmentation results plays a crucial role in ensuring both the continuity and accuracy of vessel shapes. Drawing inspiration from previous work \cite{zhang2022progressive}, we first calculate the ground truth centerline \( L_{gt} \) alongside the predicted centerline \( L_{pred} \), utilizing the ground truth vascular mask \( M_{gt} \) and the predicted mask \( M_{pred} \). To quantify the alignment between these centerlines, we compute the Dice loss, which effectively measures the overlap between \( L_{gt} \) and \( L_{pred} \). This Dice loss is then integrated into the overall model loss function, thereby enhancing the continuity and fidelity of the segmentation results. This constraint not only improves the precision of the vessel representation but also reinforces the structural coherence of the vascular network.
\begin{equation}
L_{\text{cl}} = 1-\frac{2 \times |L_{\text{gt}} \cap L_{\text{pred}}|}{|L_{\text{gt}}| + |L_{\text{pred}}|}.
\end{equation}
\subsubsection{Mask Constraint}
The Dice coefficient, a prominent evaluation metric in image segmentation, is incorporated into the PASC-Net framework as a mask constraint to enhance the continuity of vessel segmentation. This coefficient evaluates the similarity between two sets by calculating the ratio of their overlap to the total volume, providing a clear indication of segmentation performance. It is mathematically defined as follows:
\begin{equation}
L_{\text{mask}} = 1 - \frac{2 \times |M_{\text{gt}} \cap M_{\text{pred}}|}{|M_{\text{gt}}| + |M_{\text{pred}}|},
\end{equation}  
Where $M_{\text{gt}}$ represents the groundtruth of vessel segmentation, $M_{\text{pred}}$ represents the predicted result of vessel segmentation, and $L_{\text{mask}}$ represents the loss between the predicted result and the true result.

\subsubsection{Neighbor Connectivity Constraint}
Inspired by the convolutions in SSL module, we designed a Neighbor Connection (NC) prediction unit for each pixel, considering the neighboring points in eight directions around a pixel. Thus, the neighboring connection constraint collaborates with SSL Convolution to capture vascular features more effectively. Given a pixel \( P_{i,j} \), its neighboring points are defined as:
\begin{equation} 
\begin{aligned}
N_{i,j} =  \{&P_{i-1,j-1}, P_{i-1,j}, P_{i-1,j+1}, P_{i,j-1}, P_{i,j}, \\ &P_{i,j+1}, P_{i+1,j-1}, P_{i+1,j}, P_{i+1,j+1}\}, 
\end{aligned}
\end{equation}
where \( P\) represent the neighbors in eight directions of \( P_{i,j} \), respectively. Therefore, the neighboring connectivity matrix \( C \) can be defined as follows:
\begin{equation}
C_{i,j,k} = 
    \begin{cases} 
    1, & \text{if } P_{i,j} = N_k \\ 
    0, & \text{otherwise} ,
    \end{cases}
\end{equation}
where \( N_k \) represents the \( k \)-th neighbor in the neighboring point set \( N_{i,j} \), with \( k \) ranging from 1 to 8. During training, the NC branch generates 8-dimensional prediction vector for each pixel, forming 8-dimensional probability map \( C_{pred} \) that quantifies the likelihood of connections, as shown in the bottom-right of Fig.\ref{fig:spect}. Finally, the model is constrained by calculating the loss between the ground truth \( C_{gt} \) and the predicted \( C_{pred} \), which is expressed as follows:
\begin{equation}
{L}_\text{con} = -\frac{1}{N} \sum_i ( y_{i} \cdot \log(\hat{y}_i) + (1 - y_{i}) \cdot \log(1 - \hat{y}_i)),
\label{con:inventoryflow}
\end{equation}
where \( y_i \) and \( y'_i \) represent the values at corresponding positions in \( C_{gt} \) and \( C_{pred} \). 

These constraints collectively constitute the HTC module, which ultimately improves the continuity of the segmentation results.
Finally, the overall loss function \(L\) is represented as:
\begin{equation}
L = L_{\text{dice}} + L_{\text{cl}} +  {L}_\text{con} .
\end{equation}

By minimizing this loss, we aim to refine the segmentation outputs, ensuring that the predicted vascular masks closely align with the ground truth masks.

\begin{table*}[ht]
\caption{Quantitative comparison with the state-of-the-art methods on two public datasets}
\label{com}
\begin{center}
\renewcommand\arraystretch{1.4}
\resizebox{\linewidth}{!}{
\begin{tabular}{l@{\hspace{30pt}}l@{\hspace{30pt}}l@{\hspace{30pt}}c@{\hspace{30pt}}c@{\hspace{30pt}}c@{\hspace{30pt}}c@{\hspace{30pt}}c@{\hspace{30pt}}c}
\toprule

Dataset  &Method      &Publication     & Dice(\%)$\uparrow$    &  IoU(\%)$\uparrow$  & Precision(\%)$\uparrow$ & Recall(\%)$\uparrow$ & clDice(\%)$\uparrow$& ASD$\downarrow$ \\  
\midrule
\multirow{9}{*}{Arcade}
&UNet\cite{ronneberger2015u}   & MICCAI 2015 & 72.61\scriptsize$\pm$0.74& 58.37\scriptsize$\pm$1.03& 73.06\scriptsize$\pm$1.05& 74.46\scriptsize$\pm$1.01&71.20\scriptsize$\pm$0.82& 11.88\scriptsize$\pm$0.78\\
\addlinespace[1ex]
&clDice\cite{shit2021cldice} &CVPR 2021 & 73.31\scriptsize$\pm$1.09& 59.73\scriptsize$\pm$1.31& 74.13\scriptsize$\pm$1.25& 77.31\scriptsize$\pm$1.47&75.81\scriptsize$\pm$1.16& 10.76\scriptsize$\pm$0.80\\
\addlinespace[1ex]
&ResUNet\cite{xiao2018weighted} &ITME 2018 & 73.39\scriptsize$\pm$0.81& 59.45\scriptsize$\pm$1.11& 76.15\scriptsize$\pm$0.95& 73.31\scriptsize$\pm$1.19&72.70\scriptsize$\pm$0.89& 11.20\scriptsize$\pm$0.62\\
\addlinespace[1ex]
&$\mathrm{CS}^{2}$-Net\cite{mou2021cs2} &MedIA 2021 & 76.99\scriptsize$\pm$0.53& 63.67\scriptsize$\pm$0.86& 77.69\scriptsize$\pm$0.90& 78.22\scriptsize$\pm$0.68&76.68\scriptsize$\pm$0.66& 9.82\scriptsize$\pm$0.47\\
\addlinespace[1ex]
&DSCNet\cite{qi2023dynamic}&ICCV 2023   & 77.48\scriptsize$\pm$0.55& 64.36\scriptsize$\pm$0.88& 78.79\scriptsize$\pm$0.95& 77.68\scriptsize$\pm$0.58&77.47\scriptsize$\pm$0.64& 9.36\scriptsize$\pm$0.52\\
\addlinespace[1ex]
&CoANet\cite{mei2021coanet} &TIP 2021  & 80.71\scriptsize$\pm$0.36& 68.47\scriptsize$\pm$0.65& 80.65\scriptsize$\pm$0.79& 82.53\scriptsize$\pm$0.43&81.79\scriptsize$\pm$0.52& 7.43\scriptsize$\pm$0.47\\
\addlinespace[1ex]
&U-Mamba\cite{ma2024u}       &arXiv 2024            & 80.00\scriptsize$\pm$0.56&67.84\scriptsize$\pm$0.89&80.01\scriptsize$\pm$0.79&81.61\scriptsize$\pm$0.77& 81.39\scriptsize$\pm$0.69&7.87\scriptsize$\pm$0.69\\
\addlinespace[1ex]
&nnUNet\cite{isensee2021nnu}   &Nat. Methods 2021      & 81.10\scriptsize$\pm$0.57& 69.42\scriptsize$\pm$0.93& 80.51\scriptsize$\pm$0.90& 83.24\scriptsize$\pm$0.62& 82.05\scriptsize$\pm$0.77& 7.52\scriptsize$\pm$0.62\\
\addlinespace[1ex]
&\textbf{Ours}           & ---------  & \textbf{82.39}\scriptsize$\pm$0.55& \textbf{71.21}\scriptsize$\pm$0.88& \textbf{82.76}\scriptsize$\pm$0.83& \textbf{83.54}\scriptsize$\pm$0.69& \textbf{83.50}\scriptsize$\pm$0.72& \textbf{6.77}\scriptsize$\pm$0.54\\
\midrule
\multirow{9}{*}{FIVES}
&UNet\cite{ronneberger2015u}   & MICCAI 2015          & 80.80\scriptsize$\pm$1.34& 70.26\scriptsize$\pm$1.68& \textbf{94.35}\scriptsize$\pm$0.17& 73.56\scriptsize$\pm$1.92&81.77\scriptsize$\pm$1.29& 11.75\scriptsize$\pm$5.72\\
\addlinespace[1ex]
&clDice\cite{shit2021cldice}   &CVPR 2021          & 85.92\scriptsize$\pm$0.76& 76.84\scriptsize$\pm$1.04& 92.44\scriptsize$\pm$0.28& 81.71\scriptsize$\pm$1.15&87.30\scriptsize$\pm$0.75& 8.29\scriptsize$\pm$3.06\\
\addlinespace[1ex]
&ResUNet\cite{xiao2018weighted}  &ITME 2018        & 84.62\scriptsize$\pm$0.92& 75.15\scriptsize$\pm$1.24& 93.40\scriptsize$\pm$0.25& 79.41\scriptsize$\pm$1.43&86.91\scriptsize$\pm$0.88& 8.37\scriptsize$\pm$2.86\\
\addlinespace[1ex]
&$\mathrm{CS}^{2}$-Net\cite{mou2021cs2} &MedIA 2021 & 88.59\scriptsize$\pm$0.55         & 80.71\scriptsize$\pm$0.84        & 93.90\scriptsize$\pm$0.21& 84.49\scriptsize$\pm$0.79&88.62\scriptsize$\pm$0.54& 5.90\scriptsize$\pm$0.71\\
\addlinespace[1ex]
&DSCNet\cite{qi2023dynamic}    &ICCV 2023          & 88.57\scriptsize$\pm$0.53         & 80.61\scriptsize$\pm$0.79        & 92.53\scriptsize$\pm$0.22& 85.53\scriptsize$\pm$0.72&88.70\scriptsize$\pm$0.53& 6.43\scriptsize$\pm$1.20\\
\addlinespace[1ex]
&CoANet\cite{mei2021coanet}   &TIP 2021             & 89.66\scriptsize$\pm$0.50         & 82.30\scriptsize$\pm$0.71        & 94.11\scriptsize$\pm$0.12& 86.27\scriptsize$\pm$0.67&89.76\scriptsize$\pm$0.49& 5.47\scriptsize$\pm$0.87\\
\addlinespace[1ex]
&U-Mamba\cite{ma2024u}         &arXiv 2024           & 89.39\scriptsize$\pm$0.50         & 81.86\scriptsize$\pm$0.71        & 90.18\scriptsize$\pm$0.18& 89.28\scriptsize$\pm$0.67&89.67\scriptsize$\pm$0.52& 6.25\scriptsize$\pm$0.81\\
\addlinespace[1ex]
&nnUNet\cite{isensee2021nnu}    &Nat. Methods 2021         & 89.40\scriptsize$\pm$0.49         & 81.87\scriptsize$\pm$0.71        & 91.09\scriptsize$\pm$0.21& 88.30\scriptsize$\pm$0.68&89.20\scriptsize$\pm$0.53& 7.47\scriptsize$\pm$1.33\\
\addlinespace[1ex]
&\textbf{Ours}                &  ---------         & \textbf{91.83}\scriptsize$\pm$0.43&\textbf{85.83}\scriptsize$\pm$0.63& 93.44\scriptsize$\pm$0.12& \textbf{90.97}\scriptsize$\pm$0.55&\textbf{91.74}\scriptsize$\pm$0.42& \textbf{4.69}\scriptsize$\pm$0.66\\

\bottomrule

\end{tabular}}
\end{center}
\vspace{-20pt}
\end{table*}

\section{Evaluation}

\subsection{Dataset}
\subsubsection{Arcade Dataset}
The ARCADE dataset\cite{maxim2023arcade} originates from the MICCAI 2023 Coronary Artery Segmentation Challenge, comprising 1,200 coronary X-ray angiography images with a resolution of 512$\times$512 pixels. According to the official division criteria, it contains 1,000 training images and 200 test images. All images were annotated by experts using the Computer Vision Annotation Tool (CVAT). The data was collected from the Almaty Institute of Cardiology and Internal Diseases in Kazakhstan. The patient cohort had an average age of 45.8 years (median age: 60.0), with a gender distribution of 57\% male (age range: 21-85) and 43\% female (age range: 19-90).
\subsubsection{FIVES Dataset}
The FIVES dataset\cite{jin2022fives} was constructed by the Ophthalmology Center of the Second Affiliated Hospital of Zhejiang University School of Medicine. It comprises 800 high-resolution color fundus images (2048$\times$2048 pixels) from 573 patients aged 4 weeks to 83 years. According to the official division criteria, this dataset contains 600 training images and 200 test images. All images underwent rigorous quality control and were manually annotated at the pixel level by a professionally trained labeling team (including 3 senior annotators and 24 junior annotators). All annotators passed guideline-based systematic training and annotation competency assessments.

\subsection{Experimental Setup \& Evaluation Metrics}
We conducted all experiments using the PyTorch framework and trained the models on an NVIDIA RTX 3090 GPU (24GB). The nnUNet was used as the network framework for the proposed method. During network training, the Adam optimizer was employed, with an initial learning rate set to 0.01 and a batch size of 2. Due to the 2048$\times$2048 resolution of the images in the FIVES dataset, we cropped the images into 512$\times$512 patches for input into the network. During the testing phase, we made predictions for each of the four patches of an image separately and then stitched the prediction results together as the final segmentation output. 

In order to measure the performance of segmentation model, we need to choose the suitable evaluation metrics. Evaluation metrics for medical image segmentation are generally categorized into two types: overlap-based metrics and topology-based metrics. Overlap-based metrics (such as the Dice coefficient) are used to assess the degree of overlap between the segmentation result and the ground truth annotation\cite{crum2006generalized,rohlfing2011image}. Topology-based metrics (such as the Average Surface Distance and clDice) are employed to evaluate the similarity in the topological structure between the segmentation result and the ground truth\cite{shit2021cldice}. In vascular segmentation, topological structure correctness is more important than overlap, as the correct topological structure of the vessels is crucial for understanding their pathological and functional state. Therefore, combining these two metrics can more comprehensively evaluate the quality of vascular segmentation, considering both degree of overlap and topological structure consistency.

In our experiments, we used widely accepted evaluation metrics to compare the performance of PASC-Net with traditional methods, including the Dice coefficient, Intersection over Union (IoU), Accuracy, Recall, Average Surface Distance (ASD) and clDice.

\begin{figure*}[ht]
  \centering
  \includegraphics[scale=1.2]{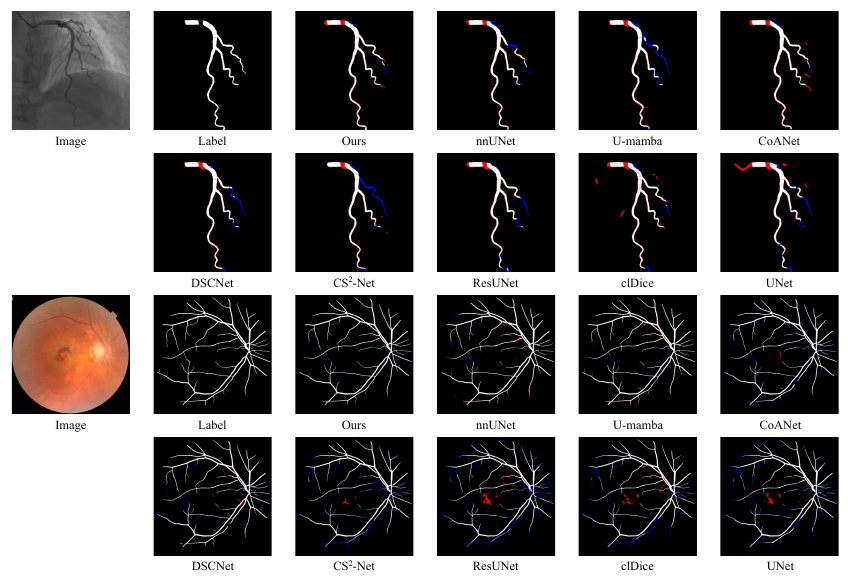}
  \caption{Qualitative comparison of our PASC-Net with other state-of-the-art methods on two public datasets. The red part of the prediction results indicates true negative (over-segmentation), and the blue part indicates false positive (mis-segmentation).}
  \label{result}
  \vspace{-20pt}
\end{figure*}

\subsection{Quantitative Experiments}
To demonstrate the superiority of our proposed framework, we compare our framework with the state-of-art methods, including UNet\cite{ronneberger2015u}, clDice\cite{shit2021cldice}, ResUNet\cite{xiao2018weighted}, $\mathrm{CS}^{2}$-Net\cite{mou2021cs2}, DSCNet\cite{qi2023dynamic}, CoANet\cite{mei2021coanet}, U-Mamba\cite{ma2024u}, and nnUNet\cite{isensee2021nnu}.

\subsubsection{Arcade dataset}
On the Arcade dataset, we compare the CNN-based approach and the topological constraint-based approach on the overlap-based metrics and topology-based metrics. The method based on topological constraints exhibits better topological metrics, with the ASD score of clDice being 0.44 lower than that of ResUNet, despite the former having a lower dice score. In addition, special models designed according to blood vessels such as $\mathrm{CS}^{2}$-Net, DSCNet, and CoANet achieved Dice scores of 76.99\%, 77.48\%, and 80.71\%, respectively. And finally, as shown in Table.\ref{com}, it can be observed that our PASC-Net outperforms other methods in all metrics. Notably, it achieves outstanding Dice and IoU scores of 82.39\% and 71.21\%, respectively, surpassing the nnUnet by margins of 1.29\% and 1.79\%. Regarding topology-aware metrics, clDice and ASD achieve superior performance compared to nnUNet, with relative improvements of 1.45\% and 0.75 respectively.

\subsubsection{Fives dataset}
On the FIVES dataset, we also compare CNN-based methods and topological constraint-based methods. The method based on topological constraints outperforms the CNN-based method on topology-base metrics, with the ASD score of $\mathrm{CS}^{2}$-Net being 1.57 lower than that of nnUnet, despite nnUnet achieving a higher dice score. In addition, special models designed according to blood vessels, such as $\mathrm{CS}^{2}$-Net, DSCNet, and CoANet, reached a similar level in Dice scores, 88.59\%, 88.57\%, and 89.66\%, respectively. And finally, the proposed method achieved an impressive Dice score of 91.83\% in the experiments, outperforming the second-best model, CoANet, by a margin of 2.17\%, as shown in Table.\ref{com}. 

Although our results are slightly lower than U-Net in terms of precision (94.35\%, a difference of 0.91\%), we achieved a significantly higher recall score, exceeding U-Net by 17.32\%. 
This indicates that our method is better at identifying more vascular structures, particularly smaller or harder-to-detect fine branches.
This is attributed to the enhanced capacity of our method to accurately capture the intricate feature of vascular tree-like tubular structure and ensure topological coherence in the segmentation results, thereby enhancing the segmentation quality.

\subsection{Qualitative Experiments}
In addition to conducting quantitative assessments, we also perform visual qualitative evaluations on two challenging samples.
\subsubsection{Arcade dataset}
The observations from Fig. \ref{result} suggest that when coronary artery is influenced by the presence of other types of vessels, all competing methods exhibit over-segmentation (\textit{e.g.}, red part), while our method closely adheres to the ground truth.
Additionally, in scenarios featuring numerous small branches within the coronary artery, most competing methods tend to exhibit segmentation discontinuity (\textit{e.g.}, blue part). In contrast, our PASC-Net, guided by supplementary hierarchical topological constraints, consistently maintains high-quality topological correctness. 
What's more, for the missing labels in the Label (possibly due to the mistakes of the labels), our method can supplement these discontinuous blood vessels.
\subsubsection{Fives dataset}
In the task of retinal vessel segmentation, our framework effectively filters out background noise in the images, allowing for a more precise focus on the vascular regions. 
{\begin{table*}[!t]
\caption{Quantitative results of ablation analysis of different components on two public datasets}
\label{abl}
\begin{center}
\resizebox{\linewidth}{!}{
\setlength{\extrarowheight}{8pt}
\renewcommand{\arraystretch}{0.76}
\begin{tabular}{|c|c|ccccc|cccccc|}
\hline

Dataset & Backbone &square&$xy$&$zw$&NC&CL  & Dice(\%)$\uparrow$   &  IoU(\%)$\uparrow$  & Precision(\%)$\uparrow$ &Recall(\%) $\uparrow$     &clDice(\%)$\uparrow$& ASD$\downarrow$\\  
\hline

\multirow{24}{*}{Arcade}
&\multirow{6}{*}{UNet\cite{ronneberger2015u}}
&\checkmark&          &          &          &          &72.61\scriptsize$\pm$0.74& 58.37\scriptsize$\pm$1.03& 73.06\scriptsize$\pm$1.05& 74.46\scriptsize$\pm$1.01&71.20\scriptsize$\pm$0.82& 11.88\scriptsize$\pm$0.78\\

& &          &\checkmark&          &          &          &76.96\scriptsize$\pm$0.55& 63.65\scriptsize$\pm$0.84& 76.07\scriptsize$\pm$0.85& 79.94\scriptsize$\pm$0.73&76.29\scriptsize$\pm$0.62& 9.72\scriptsize$\pm$0.52\\

& &          &          &\checkmark&          &          &75.76\scriptsize$\pm$0.57& 62.10\scriptsize$\pm$0.87& 76.28\scriptsize$\pm$0.84& 77.32\scriptsize$\pm$0.87&73.57\scriptsize$\pm$0.66& 9.40\scriptsize$\pm$0.45\\

& &          &\checkmark&\checkmark&          &          &77.79\scriptsize$\pm$0.54 & 64.76\scriptsize$\pm$0.85 & 76.91\scriptsize$\pm$0.84 & 80.43\scriptsize$\pm$0.70 &77.65\scriptsize$\pm$0.64 & 9.14\scriptsize$\pm$0.48\\
 
& &          &\checkmark&\checkmark&\checkmark&          &77.88\scriptsize$\pm$0.53 & 64.87\scriptsize$\pm$0.8 5& 77.20\scriptsize$\pm$0.83 &\textbf{80.47}\scriptsize$\pm$0.71  &76.94\scriptsize$\pm$0.65 &8.93\scriptsize$\pm$0.50\\

& &          &\checkmark&\checkmark&\checkmark&\checkmark&\textbf{78.00}\scriptsize$\pm$0.60 &\textbf{65.16}\scriptsize$\pm$0.94 &\textbf{80.28}\scriptsize$\pm$0.79 & 77.62\scriptsize$\pm$0.90  &\textbf{78.57}\scriptsize$\pm$0.74 &\textbf{7.82}\scriptsize$\pm$0.44\\
\cline{2-13}
&\multirow{6}{*}{FCN\cite{long2015fully}}
&\checkmark&          &          &          &          &67.24\scriptsize$\pm$0.81 &51.96\scriptsize$\pm$0.95 &70.83\scriptsize$\pm$0.93 &66.13\scriptsize$\pm$1.19  &68.11\scriptsize$\pm$0.82 &13.14\scriptsize$\pm$0.53\\

& &          &\checkmark&          &          &          &73.34\scriptsize$\pm$0.64 & 59.06\scriptsize$\pm$0.86 & 70.04\scriptsize$\pm$0.90 & 78.88\scriptsize$\pm$0.80  & 73.15\scriptsize$\pm$0.69 & 10.57\scriptsize$\pm$0.52\\

& &          &          &\checkmark&          &          &73.19\scriptsize$\pm$0.58& 58.80\scriptsize$\pm$0.81& 71.90\scriptsize$\pm$0.85& 76.48\scriptsize$\pm$0.83& 73.12\scriptsize$\pm$0.69 & 10.10\scriptsize$\pm$0.55\\

& &          &\checkmark&\checkmark&          &          &73.88\scriptsize$\pm$0.68& 59.84\scriptsize$\pm$0.95& 74.11\scriptsize$\pm$0.88& 75.30\scriptsize$\pm$0.91&73.14\scriptsize$\pm$0.79 & 10.31\scriptsize$\pm$0.60\\

& &          &\checkmark&\checkmark&\checkmark&          &74.78\scriptsize$\pm$0.63& 60.89\scriptsize$\pm$0.89& \textbf{73.50}\scriptsize$\pm$0.90& 77.86\scriptsize$\pm$0.81&73.67\scriptsize$\pm$0.70 & 9.91\scriptsize$\pm$0.50\\
& &          &\checkmark&\checkmark&\checkmark&\checkmark&\textbf{75.19}\scriptsize$\pm$0.51& \textbf{61.22}\scriptsize$\pm$0.74& 71.79\scriptsize$\pm$0.85& \textbf{80.76}\scriptsize$\pm$0.54&\textbf{78.84}\scriptsize$\pm$0.61 & \textbf{9.54}\scriptsize$\pm$0.48\\

\cline{2-13}
&\multirow{6}{*}{U-Mamba\cite{ma2024u}}
&\checkmark&          &          &          &          &80.00\scriptsize$\pm$0.56&67.84\scriptsize$\pm$0.89&80.01\scriptsize$\pm$0.79&81.61\scriptsize$\pm$0.77& 81.39\scriptsize$\pm$0.69&7.87\scriptsize$\pm$0.69\\
& &          &\checkmark&          &          &          &80.47\scriptsize$\pm$0.62&68.60\scriptsize$\pm$0.96&80.33\scriptsize$\pm$0.94&82.13\scriptsize$\pm$0.70& 80.92\scriptsize$\pm$0.79&7.70\scriptsize$\pm$0.59\\
& &          &          &\checkmark&          &          &80.24\scriptsize$\pm$0.60&68.26\scriptsize$0.96\pm$&77.70\scriptsize$\pm$0.99&\textbf{84.75}\scriptsize$\pm$0.60& 80.86\scriptsize$\pm$0.80&8.09\scriptsize$\pm$0.77\\
& &          &\checkmark&\checkmark&          &          &80.85\scriptsize$\pm$0.55&69.01\scriptsize$\pm$0.87&80.29\scriptsize$\pm$0.88&83.02\scriptsize$\pm$0.65& 81.67\scriptsize$\pm$0.77 &7.83\scriptsize$\pm$0.82\\ 
& &          &\checkmark&\checkmark&\checkmark&          &80.98\scriptsize$\pm$0.68&\textbf{69.44}\scriptsize$\pm$1.05&\textbf{81.15}\scriptsize$\pm$1.03&82.17\scriptsize$\pm$0.70& 81.66\scriptsize$\pm$0.87 &7.92\scriptsize$\pm$0.78\\
& & &\checkmark&\checkmark&\checkmark&\checkmark&\textbf{81.10}\scriptsize$\pm$0.48&69.25\scriptsize$\pm$0.81&80.79\scriptsize$\pm$0.80&83.03\scriptsize$\pm$0.57& \textbf{81.91}\scriptsize$\pm$0.67 &\textbf{7.26}\scriptsize$\pm$0.54\\
\cline{2-13}
&\multirow{6}{*}{nnUNet\cite{isensee2021nnu}}
&\checkmark&          &          &          &          &81.10\scriptsize$\pm$0.57& 69.42\scriptsize$\pm$0.93& 80.51\scriptsize$\pm$0.90& 83.24\scriptsize$\pm$0.62& 82.05\scriptsize$\pm$0.77 & 7.52\scriptsize$\pm$0.62\\

& &          &\checkmark&          &          &          &81.37\scriptsize$\pm$0.56& 69.76\scriptsize$\pm$0.88& 81.24\scriptsize$\pm$0.91& 83.11\scriptsize$\pm$0.58& 82.42\scriptsize$\pm$0.71 & 7.19\scriptsize$\pm$0.67
\\
& &          &          &\checkmark&          &          &81.71\scriptsize$\pm$0.51& 70.18\scriptsize$\pm$0.85& 82.74\scriptsize$\pm$0.83& 82.29\scriptsize$\pm$0.62& 82.45\scriptsize$\pm$0.68 & 6.82\scriptsize$\pm$0.46\\

& &          &\checkmark&\checkmark&          &          &81.95\scriptsize$\pm$0.50& 70.50\scriptsize$\pm$0.84& 82.76\scriptsize$\pm$0.76& 82.81\scriptsize$\pm$0.68& 83.12\scriptsize$\pm$0.67 & 6.71\scriptsize$\pm$0.51\\

& &          &\checkmark&\checkmark&\checkmark&          &82.35\scriptsize$\pm$0.43& 70.97\scriptsize$\pm$0.77& 82.63\scriptsize$\pm$0.77& \textbf{83.65}\scriptsize$\pm$0.51& \textbf{83.66}\scriptsize$\pm$0.58 & \textbf{6.40}\scriptsize$\pm$0.43\\

& &          &\checkmark&\checkmark&\checkmark&\checkmark&\textbf{82.39}\scriptsize$\pm$0.55& \textbf{71.21}\scriptsize$\pm$0.88& \textbf{82.76}\scriptsize$\pm$0.83& 83.54\scriptsize$\pm$0.69& 83.50\scriptsize$\pm$0.72 & 6.77\scriptsize$\pm$0.54\\
\hline
\multirow{24}{*}{FIVES}
&\multirow{6}{*}{UNet\cite{ronneberger2015u}}
&\checkmark&          &          &          &          &80.80\scriptsize$\pm$1.34& 70.26\scriptsize$\pm$1.68& 94.35\scriptsize$\pm$0.17& 73.56\scriptsize$\pm$1.92&81.77\scriptsize$\pm$1.29 & 11.75\scriptsize$\pm$5.72\\

&&          &\checkmark&          &          &          &87.22\scriptsize$\pm$0.73& 78.82\scriptsize$\pm$1.00& 93.64\scriptsize$\pm$0.27& 82.90\scriptsize$\pm$1.07&87.41\scriptsize$\pm$0.75 & 8.65\scriptsize$\pm$5.29\\

&&          &          &\checkmark&          &          &86.61\scriptsize$\pm$0.67& 77.73\scriptsize$\pm$0.92& \textbf{95.63}\scriptsize$\pm$0.16& 80.27\scriptsize$\pm$0.94&87.26\scriptsize$\pm$0.67 & \textbf{6.94}\scriptsize$\pm$1.43\\
&&          &\checkmark&\checkmark&          &          &87.45\scriptsize$\pm$0.64& 79.02\scriptsize$\pm$0.90& 93.84\scriptsize$\pm$0.11& 83.11\scriptsize$\pm$0.93&87.75\scriptsize$\pm$0.65 & 7.16\scriptsize$\pm$1.69\\

&&          &\checkmark&\checkmark&\checkmark&          &87.86\scriptsize$\pm$0.61& 79.60\scriptsize$\pm$0.85& 93.95\scriptsize$\pm$0.16& 83.50\scriptsize$\pm$0.86&88.13\scriptsize$\pm$0.61 & 7.30\scriptsize$\pm$2.22\\

&&          &\checkmark&\checkmark&\checkmark&\checkmark&\textbf{87.87}\scriptsize$\pm$0.64& \textbf{79.68}\scriptsize$\pm$0.89& 93.88\scriptsize$\pm$0.18& \textbf{83.65}\scriptsize$\pm$0.91&\textbf{88.19}\scriptsize$\pm$0.65 & 7.11\scriptsize$\pm$2.50\\

\cline{2-13}
&\multirow{6}{*}{FCN\cite{long2015fully}}
&\checkmark&          &          &          &          &82.43\scriptsize$\pm$0.63& 71.30\scriptsize$\pm$0.82& 86.12\scriptsize$\pm$0.31& 80.57\scriptsize$\pm$1.02&82.45\scriptsize$\pm$0.65 & 8.91\scriptsize$\pm$1.79\\
&&          &\checkmark&          &          &          &84.97\scriptsize$\pm$0.51& 74.89\scriptsize$\pm$0.69& 87.66\scriptsize$\pm$0.22& 83.36\scriptsize$\pm$0.77&84.64\scriptsize$\pm$0.55 & 7.22\scriptsize$\pm$1.08\\

&&          &          &\checkmark&          &          &83.80\scriptsize$\pm$0.50& 73.09\scriptsize$\pm$0.67& 87.88\scriptsize$\pm$0.21& 81.00\scriptsize$\pm$0.76&84.64\scriptsize$\pm$0.55 & 7.39\scriptsize$\pm$0.99\\

&&          &\checkmark&\checkmark&          &          &87.66\scriptsize$\pm$0.49& 79.07\scriptsize$\pm$0.71& 90.48\scriptsize$\pm$0.17& 85.82\scriptsize$\pm$0.73&87.30\scriptsize$\pm$0.52 & 6.05\scriptsize$\pm$0.73\\

&&          &\checkmark&\checkmark&\checkmark&          &88.18\scriptsize$\pm$0.48& 79.87\scriptsize$\pm$0.70& \textbf{91.98}\scriptsize$\pm$0.16& 85.43\scriptsize$\pm$0.69&87.83\scriptsize$\pm$0.49 & \textbf{5.82}\scriptsize$\pm$0.68\\

&&          &\checkmark&\checkmark&\checkmark&\checkmark&\textbf{89.00}\scriptsize$\pm$0.45& \textbf{81.14}\scriptsize$\pm$0.66& 90.82\scriptsize$\pm$0.19& \textbf{87.87}\scriptsize$\pm$0.63&\textbf{89.46}\scriptsize$\pm$0.46 & 5.92\scriptsize$\pm$0.94\\

\cline{2-13}
&\multirow{6}{*}{U-Mamba\cite{ma2024u}}
&\checkmark&          &          &          &          &89.39\scriptsize$\pm$0.50& 81.86\scriptsize$\pm$0.71& 90.18\scriptsize$\pm$0.18& 89.28\scriptsize$\pm$0.67&89.67\scriptsize$\pm$0.52& 6.25\scriptsize$\pm$0.81\\

&&          &\checkmark&          &          &          &89.74\scriptsize$\pm$0.47& 82.42\scriptsize$\pm$0.72& 91.44\scriptsize$\pm$0.31& 88.45\scriptsize$\pm$0.62&89.72\scriptsize$\pm$0.50& 7.22\scriptsize$\pm$1.65\\

&&          &          &\checkmark&          &          &89.45\scriptsize$\pm$0.47& 81.93\scriptsize$\pm$0.70& 92.53\scriptsize$\pm$0.25& 87.02\scriptsize$\pm$0.63&89.50\scriptsize$\pm$0.48& 7.46\scriptsize$\pm$1.92\\

&&          &\checkmark&\checkmark&          &          &90.14\scriptsize$\pm$0.43& 82.99\scriptsize$\pm$0.66& 90.32\scriptsize$\pm$0.27& \textbf{90.37}\scriptsize$\pm$0.54&90.31\scriptsize$\pm$0.44 & 7.18\scriptsize$\pm$1.62\\

&&          &\checkmark&\checkmark&\checkmark&          &90.55\scriptsize$\pm$0.46& 83.74\scriptsize$\pm$0.69& \textbf{93.02}\scriptsize$\pm$0.20& 88.70\scriptsize$\pm$0.61&90.56\scriptsize$\pm$0.47 & 6.35\scriptsize$\pm$1.59\\

&&          &\checkmark&\checkmark&\checkmark&\checkmark&\textbf{90.95}\scriptsize$\pm$0.44& \textbf{84.36}\scriptsize$\pm$0.66& 92.47\scriptsize$\pm$0.17& 89.99\scriptsize$\pm$0.57&\textbf{90.92}\scriptsize$\pm$0.43 & \textbf{5.65}\scriptsize$\pm$1.03\\

\cline{2-13}
&\multirow{6}{*}{nnUNet\cite{isensee2021nnu}}
&\checkmark&          &          &          &          &89.40\scriptsize$\pm$0.49& 81.87\scriptsize$\pm$0.71& 91.09\scriptsize$\pm$0.21& 88.30\scriptsize$\pm$0.68&89.20\scriptsize$\pm$0.53 & 7.47\scriptsize$\pm$1.33\\

&&          &\checkmark&          &          &          &90.32\scriptsize$\pm$0.47& 83.35\scriptsize$\pm$0.67& 92.76\scriptsize$\pm$0.17& 88.57\scriptsize$\pm$0.61&90.10\scriptsize$\pm$0.50& 6.86\scriptsize$\pm$1.27\\
&&          &          &\checkmark&          &          &90.42\scriptsize$\pm$0.48& 83.53\scriptsize$\pm$0.67& 91.57\scriptsize$\pm$0.15& 90.03\scriptsize$\pm$0.61&90.18\scriptsize$\pm$0.49 & 7.03\scriptsize$\pm$2.03\\
&&          &\checkmark&\checkmark&          &          &91.50\scriptsize$\pm$0.46& 85.32\scriptsize$\pm$0.65& 93.61\scriptsize$\pm$0.10& 90.25\scriptsize$\pm$0.61&91.41\scriptsize$\pm$0.45 & 5.09\scriptsize$\pm$0.85\\
&&          &\checkmark&\checkmark&\checkmark&          &91.70\scriptsize$\pm$0.42& 85.59\scriptsize$\pm$0.62&92.83\scriptsize$\pm$0.12&\textbf{91.25}\scriptsize$\pm$0.54&91.70\scriptsize$\pm$0.40 &4.74\scriptsize$\pm$0.67\\
&&          &\checkmark&\checkmark&\checkmark&\checkmark&\textbf{91.83}\scriptsize$\pm$0.43&\textbf{85.83}\scriptsize$\pm$0.63& \textbf{93.44}\scriptsize$\pm$0.12& 90.97\scriptsize$\pm$0.55&\textbf{91.74}\scriptsize$\pm$0.42 & \textbf{4.69}\scriptsize$\pm$0.66\\

\hline
\end{tabular}}
\end{center}
\end{table*}}
Compared to other segmentation methods, our framework demonstrates superior performance in addressing both over-segmentation and mis-segmentation issues. Specifically, the proportion of over-segmentation (highlighted in red) and under-segmentation (highlighted in blue) is significantly reduced in our model's segmentation results compared to others. Compared with the segmentation results of other methods, our results are obviously optimal both in the overall topology of blood vessels and in the detection of small branches of blood vessels, and are very close to the label.

\subsection{Ablation Experiments}
To further validate the effectiveness of our proposed modules, we conducted extensive ablation studies on two public datasets using four widely adopted segmentation architectures, as summarized in Tab.~\ref{abl}. Specifically, we integrated the SSL and HTC modules into U-Net, nnU-Net, FCN, and U-Mamba to evaluate their generalizability and performance improvements across different network backbones.

\subsubsection{Different configurations of SSL module}
We designed three types of SSL convolutions for different directions: $xy$-based (horizontal and vertical directions), $zw$-based (two diagonal directions), and $xyzw$-based convolutions, and compared with traditional square convolutions. The experimental results show that all three types of SSL convolutions outperformed ordinary square convolutions. Taking the results of the nnUNet architecture on the ARCADE dataset as an example, the Dice score using standard convolutions was 81.10\%. After employing $xy$-base and $zy$-base convolutions separately, the Dice scores improved to 81.37\% and 81.71\%, representing increases of 0.27\% and 0.61\%, respectively. Further integrating $xy$-base and $zw$-base ($xyzw$-base) convolutions achieved a Dice score of 81.95\%, marking an improvement of 0.85\%. Among them, the $xyzw$-based convolution achieved the best results, surpassing the $xy$ and $zw$-based convolutions. This indicates that using convolutions with offsets in four directions can more effectively capture the tubular characteristics of vessels, as it is easier to represent vessel distributions in space, particularly for complex vascular branching structures.

\subsubsection{Different components of HTC module}
Our HTC module consists of three key units: surface constraint, neighbor connectivity constraint(NC), and centerline constraint(CL). In our experiments, we first applied only the mask constraint to evaluate the segmentation performance. The fourth row in the Tab.\ref{abl} for each network architecture shows the segmentation results when only the mask constraint was used. Taking the results of the nnUNet architecture on the ARCADE dataset as an example, the plane-constrained model achieved a Dice score of 81.95\%. After sequentially incorporating the NC unit and CL unit, the Dice scores improved to 82.35\% and 82.39\%, corresponding to increases of 0.40\% and 0.44\%, respectively. The experimental results demonstrate significant improvements as these two units were incorporated, highlighting their effectiveness in improving the segmentation performance.

These experiments sequentially validated the effectiveness of the SSL module and the HTC module.

\section{Conclusion}
In this study, we propose a novel vascular segmentation framework PASC-Net. This framework combines shape self-learning convolution and hierarchical topology constraints. By embedding the shape self-learning convolution into traditional convolutional neural networks, it can more accurately perceive fine-grained features of tubular structures, significantly reducing the interference from background noise. Additionally, the introduction of the Hierarchical Topology Constraint module effectively mitigates the issue of poor topology in vascular segmentation, preserving coherence and integrity of vascular topology.
Experimental results demonstrate that PASC-Net outperforms existing methods across multiple vascular segmentation datasets. Whether based on the U-Net, FCN, U-Mamba, or nnUNet frameworks, our method significantly improves model performance. Particularly within the nnUNet framework, PASC-Net achieves state-of-the-art performance.

In future work, we plan to extend this method to 3D image segmentation to further validate its performance in more complex medical imaging tasks.

\printcredits

\bibliographystyle{cas-model2-names}

\bibliography{cas-refs}





\end{document}